# Analysis of publications by authors of Ukrainian institutes in Scopus-delisted titles

Serhii Nazarovets
serhii.nazarovets@gmail.com
http://orcid.org/0000-0002-5067-4498

*Abstract*

In Ukraine, Scopus data are used to evaluate academics. Existing shortcomings in the Ukrainian evaluation system allow them to publish in titles that have been delisted from Scopus, and continue to use those papers as credible research output for evaluation. The purpose of this study was to analyze the publishing activity of Ukrainian institutions in Scopus-delisted titles (September 2021) in different fields between 2011 and 2020 and to attempt to appreciate how common this practice is among Ukrainian authors. Scopus was sourced to collect bibliographic and citations-related data, while SciVal was used to analyze these data. The findings suggest that for 17 Ukrainian institutions, papers from titles that have been delisted from Scopus still play an important part of the publication achievement of their employees. In particular, in the field of economics, econometrics, and finance, 46.92% of Ukrainian papers were published in a title that was excluded from Scopus. Moreover, the analysis indicated that in two Ukrainian institutions, the level of citation of such papers significantly exceeds the average number of citations to Scopus-indexed papers in the same year and in the same field. Given that bibliometric indicators are also used for research assessment in other Eastern European countries, the results of this paper are applicable to a wider geographic context.

*Key points*

- Ukrainian evaluation practices allow authors to be rewarded for publications in Scopus-delisted titles, as equally as publications in more prestigious journals, amplifying the number of papers by Ukrainian researchers in such titles.

- For 17 Ukrainian institutions and universities, publications in Scopus-delisted titles accounted for a large percentage of their total number of publications, and in the field of economics, econometrics and finance, 46.92% of papers from Ukraine were published in Scopus-delisted titles.

- The citation of Ukrainian papers in Scopus-delisted titles significantly exceeds the number of citations to papers in Scopus-indexed journals in the same year and field.

*Introduction*

In Ukraine, as in many Eastern European countries, quantitative data from Scopus and the Web of Science (WoS) Core Collection are qualitative and quantitative indicators that are used at the state level to evaluate research results, to make promotion-related decisions, to distribute financial awards, and also to evaluate applications for funding of research projects (Zacharewicz et al., 2019; Sivertsen, 2020; Jappe, 2020). In the national evaluation systems of these countries, bibliometrics form part of a set of indicators that are used to distribute research funding. For instance, according to research evaluation in the Czech Republic, universities score points for papers published in Scopus-indexed journals or in journals with a non-zero Clarivate impact factor (Fiala, 2013). In the evaluation system in Romania, papers that are published in journals that are indexed in Clarivate's Journal Citation Reports (JCR) have a greater weight during assessment (Vîiu, Păunescu & Miroiu, 2016). In Slovakia, publications in journals that are indexed in WoS and Scopus databases are required to obtain an academic position (Ciaian, Lancaric & Pokrivcak, 2018). In Russia, a program to increase the competitiveness of universities pays considerable attention to publications indexed in WoS and Scopus (Guskov, Kosyakov & Selivanova, 2018).

In Poland, publications in journals indexed in JCR were previously rated higher than publications in journals not included in this list (Korytkowski & Kulczycki, 2019). However, in 2018, Poland adopted new national rules, or Law 2.0, also referred to as the Constitution for Science, which does not mention specific lists of scientific publications, but only indicates "international databases of the [sic] scientific journals with the broadest coverage" (MSHE, 2018).

Belarus is another example of an Eastern European country that does not use Scopus or WoS data directly in its national research evaluation system. Belarus uses lists of scientific publications for research evaluation, for example, in the process of awarding an academic degree (HAC, 2014). The peculiarity of these lists is that they contain many national non-English language journals. Curiously, national journals are widely represented in the Hungarian Scientific Bibliography (HSB), which plays a key role in research evaluation in Hungary (Csomós, 2020). However, this Hungarian national publication and citation database also contains data and indicators from commercial databases. In particular, journals' quartiles according to the Scimago portal, which uses Scopus data, are integrated into HSB to evaluate journals[1].

The quality of the content and the reliability of the bibliometric indicators provided by WoS and/or Scopus databases are very important for users in these countries. Today, scientific communication suffers from problems associated with so-called "predatory" journals and fake publications (Gasparyan et al., 2016). Publishers of such journals and conference proceedings are not interested in conducting quality peer review of author's manuscripts, seeking instead to draw a publication fee from as many authors as possible, regardless of the scientific quality of the manuscripts they publish (McLeod, Savage, & Simkin, 2018). There is no doubt that the owners of such suspicious journals are interested in having their journal titles indexed in WoS and Scopus databases in order to increase their number of potential clients. It is also clear that authors originating from countries where indicators from these global commercial databases are used in evaluation processes risk falling victim to such pseudo-journals, especially if the authors are under pressure by their employer to publish, and are unsure of their ability to present results in authoritative journals (Xia et al., 2015).

New titles are added to Scopus[2] and the WoS Core Collection[3] only after they have been carefully evaluated by relevant experts. In addition to evaluating journals for inclusion, these experts also examine journals that are already indexed in Scopus and WoS. If the indexed journal demonstrates unethical behavior, or if the quality of the journal's publications or its bibliometric indicators no longer meet the database's criteria, then indexation of current issues of such journals is stopped. For example, titles may cease to be indexed in Scopus for different reasons: low quantitative bibliometrics indicators of the title compared to other titles in the same subject field, concerns about the publisher's publication standards, or journal outlier behavior.

The list of journal titles that have been delisted from Scopus is regularly updated and publicly available on the Elsevier website. Unfortunately, the list of WoS-delisted titles is not publicly available, so this study is limited to an analysis of Scopus-delisted titles. In Scopus, the term "discontinued" is used to refer to titles whose newly published papers are no longer indexed in this database. However, most of these titles continue to be published, and only their indexation in Scopus has stopped. In order not to confuse readers, this study uses the adjective "delisted" to denote these titles.

[1] https://www.mtmt.hu/sites/default/files/utmutatok/szakteruleti_folyoiratrangsor_az_mtmt-ben_20161025.pdf
[2] https://www.elsevier.com/solutions/scopus/how-scopus-works/content/content-policy-and-selection
[3] https://clarivate.com/webofsciencegroup/journal-evaluation-process-and-selection-criteria/

During 2011-2020, 678 titles were delisted from Scopus. Figure 1 indicates that the Scopus experts began to actively re-evaluate and suspend the indexing of publications in 2016. Figure 2 presents the percentage of publications that have been delisted from Scopus based on subject areas from 2011-2020. Most of the Scopus-delisted titles are in the fields of engineering (34%), computer science (9%), and materials science (8%).

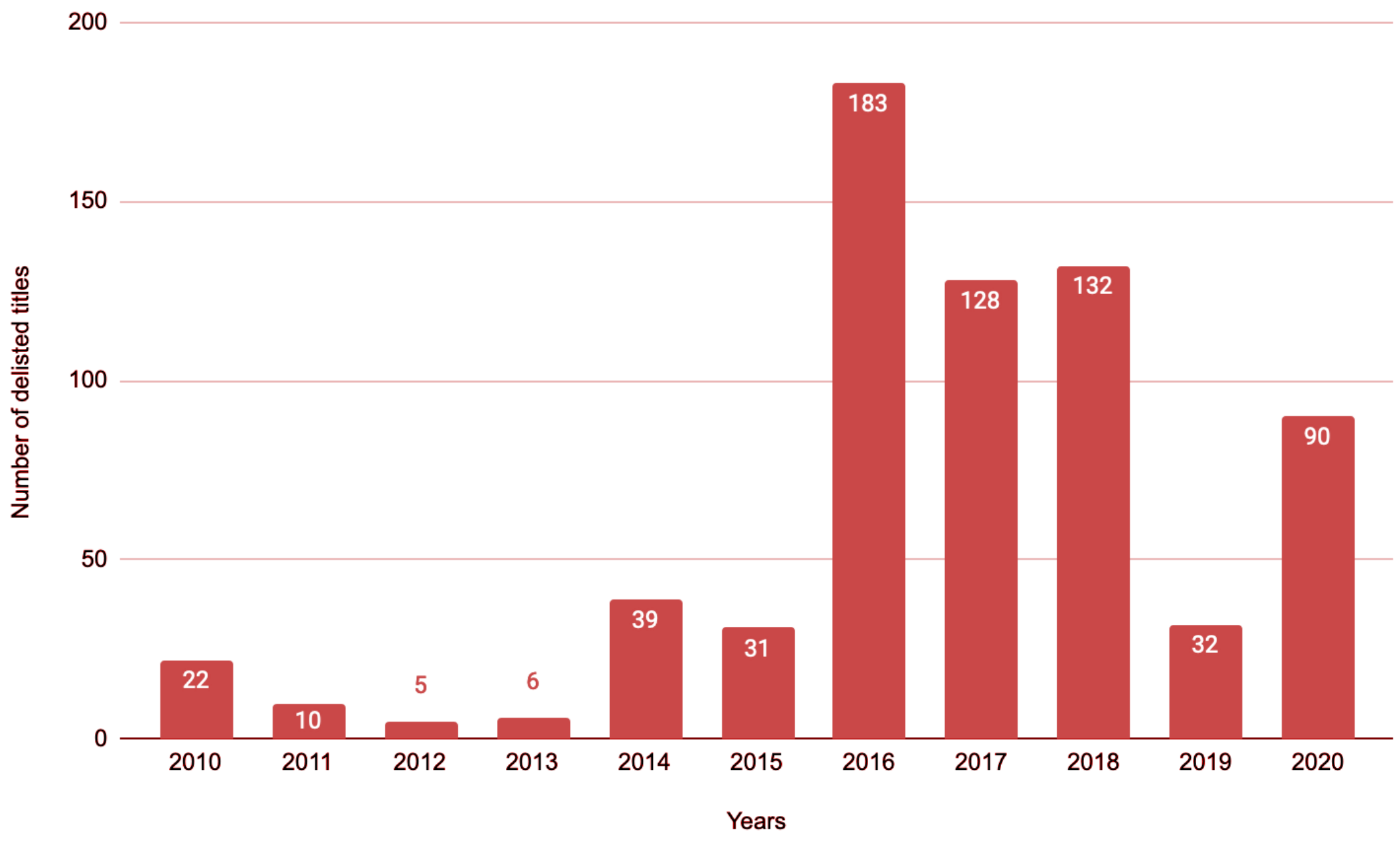

**Fig. 1** Number of Scopus-delisted titles in 2010-2020

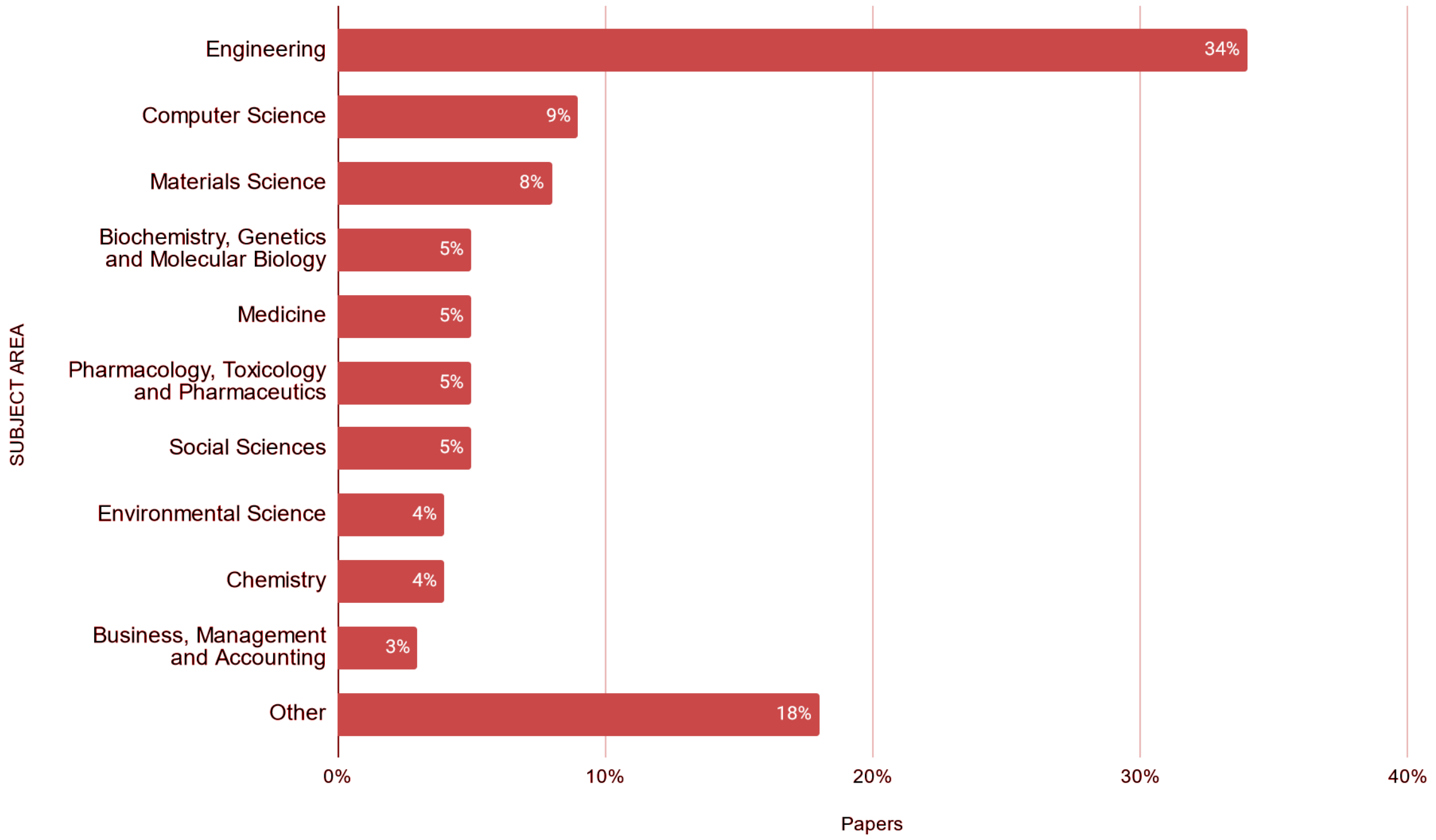

**Fig. 2** Percentage of publications in Scopus-delisted titles by subject areas in 2011-2020

However, even if current issues of a particular journal title are no longer indexed, already indexed papers of this title are not deleted from the database (except in rare cases). Thus, authors can be cited (and self-cite), even papers that have been published in delisted titles, allowing, to some extent, the bibliometrics indicators of scientists, research groups, and entire institutions to improve (Cortegiani et al., 2020; Moussa, 2021). This feature of indexing content in Scopus and WoS creates difficulties for bibliometric analyses or research evaluation because it might distort the results. The systematic inclusion of such papers and their citations in the process of evaluating research output, and equating them to quality peer-reviewed work, may have negative consequences for scientific development.

In this study, "delisted titles" are not equated with "predatory/fake titles". Previous studies claimed that potentially predatory journal titles are also indexed in Scopus (Marina & Sterligov, 2021). However, as argued by Krawczyk & Kulczycki (2021), the term "predatory journal" is quite controversial and in practice it is difficult to establish unambiguous formal criteria for identifying illegitimate scientific titles (Teixeira da Silva et al., 2022; Yamada & Teixeira da Silva, 2022). The author is of the opinion that there is a distinct difference between predatory and Scopus-delisted journal titles. Truly predatory journals are not interested in quality peer review or adherence to high publishing standards, and their primary goal is to obtain article processing charges (APCs) from authors, in the case of open access journals (Xia, 2015). The behavior of many journal titles that were later excluded from Scopus, in the first years of their indexation, did not differ considerably from other journals in the same field. Suspicious behavior, which is inherent in predatory journals, including very rapid growth in the number of papers, typifies the behavior of select journals in their more recent years of indexing in Scopus. Such a decline in quality may have occurred for a variety of reasons, and not necessarily due to the transformation of the journal into an overtly predatory title. For example, an editorial office may have started to work with so-called “paper mills”, which imitate the work of editorial services but instead offer their clients the sale of papers, citations of previous work, thereby artificially raising individual metrics, or selling authorship (Rivera & Teixeira da Silva, 2021; Else & Van Noorden, 2021; Abalkina, 2021).

Independent of the reasons, there is no doubt that the exclusion of a journal title from a database is related to a decrease in the quality and metrics of its scientific content. Therefore, to characterize delisted titles, the epithet "questionable" can be used (Eykens et al., 2019; Nelhans & Bodin, 2020).

An author who may have become the victim of aggressive marketing by a questionable journal might be unaware of that journal’s problems with peer review and/or compliance with other academic publishing standards (Cobey et al., 2019; Krauskopf, 2018). Today, many scholars are under pressure from their university administration, which requires them to constantly produce papers (Mills & Inouye, 2021). Accordingly, since some scholars worry that they will not be able to publish in authoritative journals, they may prefer to submit to questionable titles (Shaghaei et al., 2018; Mertkan, Onurkan Aliusta & Suphi, 2021). However, some scholars are consciously collaborating with predatory publishers to systematically improve their bibliometrics and receive undeserved government awards (Pond et al., 2019), some of whom may feign ignorance as a screen to justify their unethical behavior (Frandsen, 2019). Also, many universities and funders inadvertently support questionable publishers when the leaders of these universities use erroneous guidelines and are primarily interested in achieving certain quantitative indicators of scientific output (Siler et al., 2021). Therefore, the facts of publication of articles in delisted titles need to be carefully verified by experts in charge of scientific evaluation.

*Use of citation databases in Ukrainian research evaluation*

Ukraine is the largest country in Central-Eastern Europe. It regained its independence in 1991 after the collapse of the Soviet Union. However, the young Ukrainian democracy still cannot logically complete a number of important reforms, including in the fields of education and science (Hladchenko, 2020). The lack of transparency of evaluation of scientific achievements, doubts about the relevance of research productivity of many Ukrainian institutions, and frequent cases of plagiarism in scientific papers have led to the devaluation of expert evaluation in Ukraine. Ukrainian taxpayers have accumulated many questions about the transparency of the procedure for allocating public funding for research, as well as the procedure for awarding academic titles to research and teaching positions. In order to change this situation for the better, Ukrainian officials have attempted to supplement expert assessments with formal bibliometric indicators.

In Ukraine, since Soviet times, there has been a two-level system of academic degrees – Candidate of Sciences (PhD) and Doctor of Sciences (DSc) (a higher degree which may be earned after the Candidate of Sciences), as well as a two-level system of academic ranks – Docent (Associate Professor) and Professor. Previously, in order to obtain a degree in Ukraine, it was necessary to have published a certain number of articles in Ukrainian journals included in the List of Professional Publications of Ukraine, regardless of the number of articles in foreign journals. This irritated some leading Ukrainian scientists, who sought to present their results primarily in influential foreign journals, rather than in local Ukrainian journals with a restricted readership. One of the first initiatives to implement bibliometrics at the national level was the Order of the Ministry of Education, Science, Youth and Sports of Ukraine (MESU, 2012). According to MESU, publications in journals included in "international scientometric databases" were equivalent to publications in Ukrainian journals for obtaining Doctor of Sciences and Candidate of Sciences degrees. However, the text of the Order did not contain a list of these databases, and therefore applicants could, as such, consider their papers published in almost any foreign journals and conference proceedings. At the same time, most Ukrainian authors simply continued to publish mainly in Ukrainian journals (Hladchenko & Moed, 2021).

The portal of the Vernadsky National Library of Ukraine presents information on more than 2,800 Ukrainian journals[4]. In contrast, Scopus only indexes just over 100 Ukrainian journals. The vast majority of Ukrainian journals publish articles in Ukrainian, but these titles are poorly represented in reference databases, so researchers from other countries rarely use these journals. Changes in the approach to the evaluation of research in Ukraine have significantly affected the choice of communication channels of Ukrainian scientists to present research results. For example, over the past 10 years, the annual number of publications by Ukrainian scientists in Scopus-indexed titles has more than doubled, 8,400 documents in 2011 versus more than 19,700 documents in 2020. It should be noted that this increase in the number of documents was accompanied by an increase in the number of publications by Ukrainian authors in Ukrainian journals indexed in Scopus (Nazarovets, 2020). Nevertheless, it can be stated that, at least in quantitative terms, Ukrainian national initiatives to use information from citation databases to assess research output have yielded some positive results.

For the first time, a clear requirement for Ukrainian scientists to have publications in journals that are included in the WoS Core Collection or Scopus appeared in 2015 in the Resolution of the Cabinet of Ministers of Ukraine on approval of the license conditions for educational activities (Cabinet of Ministers of Ukraine, 2015a). In 2015, publications in international journals began to be taken into account for obtaining an academic rank (MESU, 2015b). Updated in 2017, the methodology for evaluating the effectiveness of scientific institutions of the National Academy of Sciences of Ukraine used the number of papers and the h-index of the scientist according to Scopus or WoS (NASU, 2017). In 2018, a resolution of the Cabinet of Ministers of Ukraine approved the procedure for conducting state certification of universities for their scientific activities, which also uses indicators from Scopus or WoS (Cabinet of Ministers of Ukraine, 2018). Also, in 2018, the procedure for forming the list of scientific professional publications of Ukraine was updated. In the new procedure, Ukrainian journals indexed in the WoS Core Collection or Scopus were automatically included in this list without any additional checks. In contrast, all other Ukrainian journals had to meet a number of formal requirements in order to be included in this list (MESU, 2018). In 2018, the National Research Foundation of Ukraine was established to conduct an open competitive selection of research projects. Bibliometric indicators of project authors are taken into account in the process of evaluating grant applications submitted to the competitions of this Foundation (NRFU, 2021). In 2019, there were updated requirements for obtaining an academic rank that included indexing in the WoS Core Collection and Scopus databases (MESU, 2019a). In addition, the order of the Ministry of Education and Science noted that an applicant must have publications in journal titles that are included in Scopus or WoS in order to obtain a rank of Associate Professor or Professor (MESU, 2019b).

[4] http://www.irbis-nbuv.gov.ua/cgi-bin/irbis_nbuv/cgiirbis_64.exe?C21COM=F&I21DBN=UJRN&P21DBN=UJRN

Thus, over the past ten years, Ukrainian scientific managers have increasingly used Scopus or WoS data to evaluate the scientific performance of scientists and institutions. It should also be noted that in 2021, the Ministry of Education and Science of Ukraine approved an updated roadmap for the integration of Ukraine's research and innovation system into the European Research Area (MESU, 2021). This roadmap states that the Ministry plans to continue to evaluate the effectiveness of scientific institutions by taking into account the number of papers of Ukrainian scientists presented in Scopus and WoS.

It is also important to mention that in January 2022, the Cabinet of Ministers of Ukraine approved a new procedure for awarding a PhD degree, according to which the applicant is required to have three articles in Ukrainian journals included in the List of Professional Publications of Ukraine. In this case, according to this procedure, an article in journals that are ranked in Q1-Q3 according to SCImago Journal Rank (SJR) or JCR is equated to two articles in journals from the List (Cabinet of Ministers of Ukraine, 2022). As of January 2022, this List contained more than 1,400 Ukrainian journal titles. This scaling-down of PhD requirements might have negative consequences, because some universities that require their academics to publish in these journals, could favor output quantity over quality (Nazarovets, 2022).

Summarizing the above, in Ukraine, data from citation databases are actively used in the process of awarding degrees, in the evaluation of journals, projects, authors, universities and research institutions. Publications in titles that are indexed in Scopus and WoS are evaluated separately from publications in non-indexed Ukrainian journals. However, in the Ukrainian evaluation system, there is no universal scale for scoring metrics of publications in different scientific fields, and there are no clear procedures for verifying these data. The evaluation and comparison of indicators every time takes place according to the objectives of a particular evaluation, and takes into account primarily the number of publications. Therefore, in the absence of other clear guidelines, many Ukrainian authors try to publish as many papers as possible in indexed journals without paying much attention to their reputation or bibliometric indicators.

*Scopus-delisted titles in research assessments*

Previous research has shown that, in response to the introduction of evaluation criteria, authors often begin to use different adaptation mechanisms. For example, a study of the impact of national evaluation on the publication strategies of sociologists in Italy showed that the authors focused on presenting their results primarily in journals, which would allow them to ensure the number of publications, regardless of the quality of these journals (Akbaritabar, Bravo & Squazzoni, 2021).

A huge disadvantage of all the above-mentioned Ukrainian legislative initiatives is that these rules reward authors for the mere publication of a paper indexed in Scopus or WoS, even if that journal ceased to be indexed due to gross violations of scientific ethics. Such remuneration includes, for example, a guaranteed lifetime surcharge to their salary for obtaining a PhD degree, as well as the opportunity to hold certain positions in universities and research institutions of Ukraine.

The applicant for an academic degree or a research grant in Ukraine simply indicates in their application a list of papers that are indexed in Scopus and/or WoS Core Collection. Applicants are expected to report their publications, honestly. In fact, the experts who evaluate such applications only collect quantitative information, without taking into account the current status of the journal indexing, number of self-citations, or other factors, in the evaluation process.

Some Ukrainian authors might not be very interested in the effective presentation of the results of their research, but instead might be more focused on the rapid formal implementation of the Ministry's requirements in order to receive appropriate reputational and financial rewards from the state. Meanwhile, some heads of research institutions might also be interested in increasing the bibliometric indicators of their employees, and thus their institute, through questionable publications, in order to pass certification and to improve the institution's chances of receiving public funding. Moreover, papers published in such academically suspect journal titles might continue to influence the distribution of research funding in Ukraine.

Despite the predictive adaptive behavior of Ukrainian authors, Ukrainian officials have not provided any tools in the requirements to prevent possible manipulation. For example, the evaluation process should have limited the impact of self-citations, publications in journal titles excluded from Scopus or WoS, and citations from these journals, and issuing penalties where citation cartels might exist. However, the Ukrainian research evaluation system has not yet introduced any measures to deal with such situations.

It is also worth noting that Ukrainian science is poorly funded, not only in comparison with developed countries, but also in comparison with Ukraine's neighboring countries of the former Soviet Union. Table 1 shows that Ukraine's Research and Development expenditure (% of GDP) has been declining in recent years and is one of the lowest in Eastern Europe (World Bank, 2020). Given the weak funding for science in Ukraine, the use of imperfect quantitative indicators for the distribution of funding may lead to the allocation, even of small funds meant for scientific development and research purposes, for other inappropriate and/or unrelated purposes, as has been shown in examples from around the world (Hedding, 2019).

Table 1 Research and development expenditure (% of GDP) in select countries of Eastern Europe in 2011-2018

Therefore, the main research questions of the study were:
(1) How many papers have Ukrainian academics published in Scopus-delisted journal titles (institutional level), and how often have these papers been cited?
(2) How does the Ukrainian publishing practice in Scopus-delisted journal titles compare with the publishing practice in other Eastern European countries?
(3) How have Ukrainian legislative initiatives influenced the publication behavior of Ukrainian authors regarding publication in Scopus-delisted journal titles in different subject areas?

Publications in Scopus-delisted journal titles are considered in this study only in the context of Ukrainian national evaluation. Therefore, in this paper, all calculations are made on the basis of papers that were indexed in Scopus. It is possible that Ukrainian authors continued to publish their papers in these titles even after they stopped being indexed in Scopus, but these papers were not taken into account in this study.

*Methodology*

The official list of Scopus-delisted titles was used in the study (last updated in September 2021). Papers were searched in the Scopus database based on the ISSN of these journal titles. These 2011-2020 papers were then screened to identify papers with authors who were employees of Ukrainian institutions. An example of a search query was: ISSN (1993-6788) AND (LIMIT-TO (AFFILCOUNTRY, "Ukraine")). Moreover, a search by Source title was used for several entries that did not contain ISSNs: SRCTITLE ("Academic Journal of Cancer Research") AND (LIMIT-TO (AFFILCOUNTRY, "Ukraine")). Data for other countries were similarly obtained. For additional analysis, three types of publications were selected: article, conference paper and review.

The search results were generated in a separate dataset (see Supplementary data). This dataset was imported for analysis in SciVal (https://www.scival.com/) using Scopus EID, which are unique academic work identifiers assigned in the Scopus database. The dataset was created, imported and analyzed on November 22, 2021. The following were analyzed: number of papers in Scopus-delisted titles and their percentage of the total number of Ukrainian papers indexed in Scopus by research fields and institutions; number of citations and number of authors of these papers; structure of co-authorship of these papers. The research field of a paper was defined according to the All Science Journal Classification (ASJC) code of the journal title in which it was published. In Scopus, an indexed journal title can have several ASJC codes simultaneously, so the same analyzed paper can belong to several research fields as well.

In addition, the average citation of papers of each institution and Field-Weighted Citation Impact (FWCI) of these papers was calculated. FWCI is an article age- and field-normalized metric to evaluate scientific visibility and impact, and is defined as the ratio of total citations received by any given article to total citations that would be expected based on the average of that particular topic in the same time span (Zanotto & Carvalho, 2021). For this study, FWCI was obtained using SciVal based on all Scopus data. The percentage of papers in Scopus-delisted journal titles relative to the total number of papers of countries and employees of Ukrainian institutions was also established.

*Results*

In 2011-2020, employees of Ukrainian institutions published 6,902 papers in Scopus-delisted journal titles (5,027 articles, 1,829 conference papers, and 46 reviews). This accounts for 6% of the total number of papers published by the same Ukrainian authors during this period. Conference proceedings were cited more often than articles or reviews - 3.7 mean citations per paper (articles - 2.4; reviews - 3.1). According to Scopus, during 2011-2020, 16,265 Ukrainian authors published papers in Scopus-delisted titles, or 18% of the total number of Ukrainian authors. Thus, almost one in five Ukrainian authors published one or more papers in Scopus-delisted journal titles.

Figure 3 shows that the number of such papers grew rapidly in the second decade of this century, and their number has continued to grow in recent years. Among all delisted journal titles, most articles were published by Ukrainian authors in the journal *Actual Problems of Economics* (1675 papers). The indexation of this journal was stopped in 2017, and might explain the sudden decrease in the number of papers in Scopus-delisted journal titles in 2017. Ranked second and third in terms of the number of papers by Ukrainian authors in Scopus-delisted journal titles were two conference proceedings: *Advances in Intelligent Systems and Computing* (685 papers in the field of computer science and engineering) and *IOP Conference Series Materials Science and Engineering* (586 papers in the field of engineering and materials science).

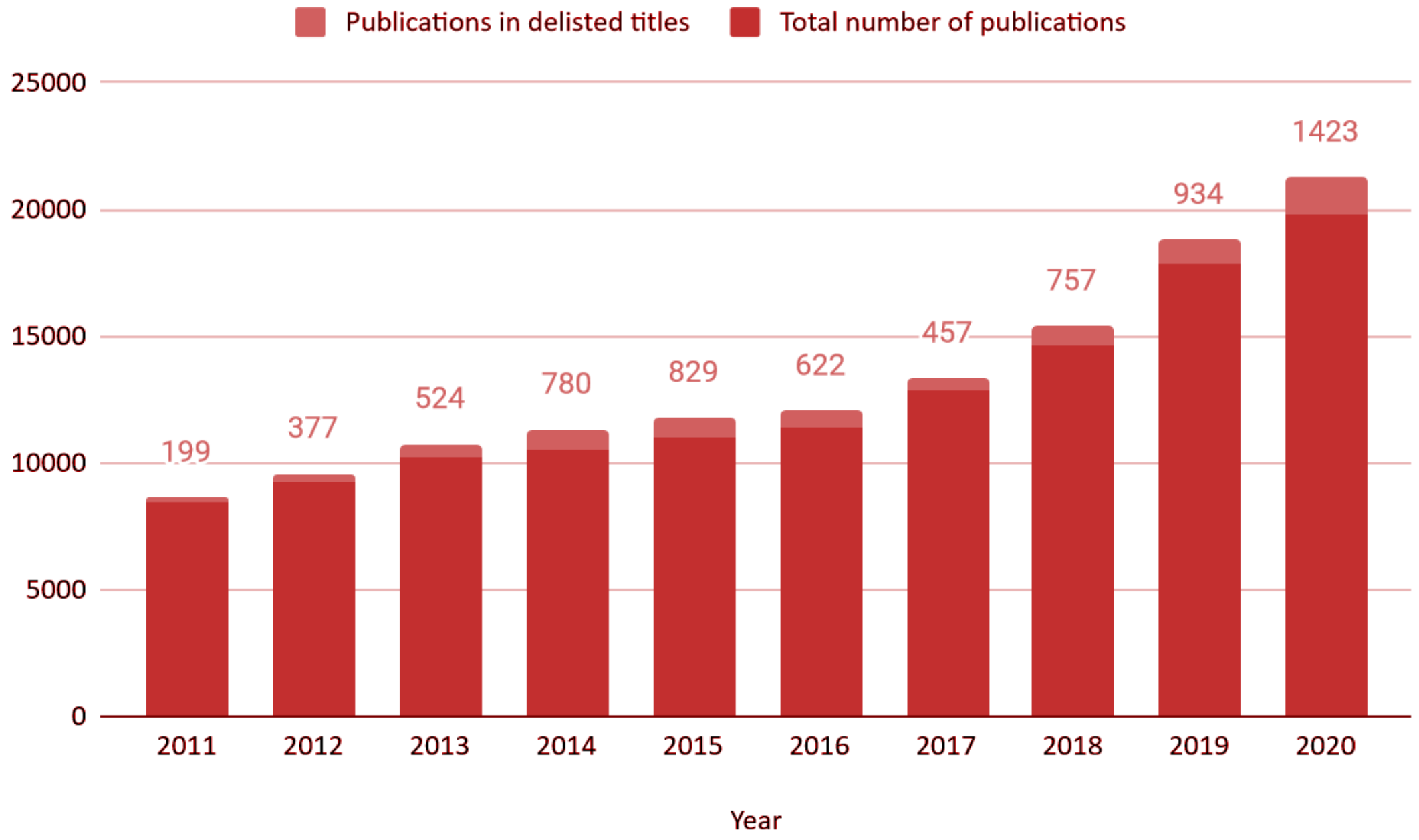

**Fig. 3** Dynamics of publishing activity of Ukrainian authors in Scopus-delisted journal titles in 2011-2020

Table 2 shows the change in the number of papers published by Ukrainian authors based on year in Scopus-delisted journal titles in which they published the largest number of papers during 2011-2020. It can be seen that after Scopus stopped indexing popular journal titles among Ukrainian authors, the number of papers in other journals began to increase sharply, even though Ukrainian authors had rarely or had not published in these titles.

**Table 2** Publication activity of Ukrainian authors in Scopus-delisted journal titles in which they published the most papers in 2011-2020

Table 3 demonstrates that most papers in Scopus-delisted journal titles by Ukrainian authors were written as an institutional collaboration (33.6%). Conversely, the fewest such papers were written as part of international cooperation (18.4%). Thus, with the participation of foreign co-authors, Ukrainian researchers were much more likely to select more reliable journal titles to present research results than when they wrote papers as part of an institutional collaboration.

**Table 3** Papers published by Ukrainian authors during 2011-2020 in Scopus-delisted journal titles based on the amount of international, national and institutional collaboration

Table 4 shows that papers published by Ukrainian scholars in Scopus-delisted journal titles represent a small percentage of the total number of Ukrainian papers in most fields. However, it is unexpectedly high for some subject fields, especially for econometrics and finance (46.92%) and pharmacology, toxicology and pharmaceutics (18.6%).

**Table 4** Papers published by Ukrainian authors during 2011-2020 in Scopus-delisted journal titles based on subject field and their percentage of the total number of Ukrainian papers in those fields

Table 5 shows an analysis of citations of papers written by Ukrainian authors in Scopus-delisted journal titles during 2011-2020. The citation rate of these papers for many Ukrainian institutions is unexpectedly high. The highest citation rate is by Ivan Kozhedub Kharkiv National Air Force University – more than 10 citations on average for each paper. Papers in Scopus-delisted journal titles of authors from the Odesa National Polytechnic University, Kharkiv National University of Radio Electronics and the Lviv Polytechnic National University received an average of six citations for each paper. The FWCI also indicates that papers from some Ukrainian universities and scientific institutions were cited much more than expected.

**Table 5** Papers published by Ukrainian authors during 2011-2020 in Scopus-delisted journal titles based on institutions and their percentage of the total number of Ukrainian papers in each field

An analysis of papers published by Ukrainian authors during 2011-2020 in Scopus-delisted journal titles shows that there are numerous institutions in Ukraine whose employees have published a significant percentage of the institution's work in delisted journal titles. The institutes that show a reverse trend of this indicator are the Dumansky Institute of Colloid and Water Chemistry (52.52%), the National Bank of Ukraine (47.12%), and Poltava National Technical Yuriy Kondratyuk University (43.33%) (Table 5).

In order to better understand the indicators of publishing activity of Ukrainian authors in Scopus-delisted journal titles, these indicators were compared with the indicators of authors from other Eastern European countries, namely Belarus, the Czech Republic, Hungary, Poland, Romania and the Russian Federation.

Table 6 shows that in general, those publications in Scopus-delisted journal titles do not represent a significant percentage of the total number of publications from those countries, ranging from the lowest percentage in Hungary (1.14%) to the highest in Romania (8.7%). However, there are differences between countries in terms of subject areas in which these papers have been published. The subject area economics, econometrics and finance is most commonly ranked at the top, in 5 out of 7 countries. Moreover, this percentage is particularly high in Ukraine (46.92%) and Russia (31.49%).

**Table 6** Number and percentage of papers in Scopus-delisted journal titles of authors from seven Eastern European countries during 2011-2020. Top-3 subject areas based on the percentage of papers in Scopus-delisted journal titles, by countries

*Conclusions*

Today, governments in many countries are trying to maximize their investment in science, so research process is increasingly being analyzed, evaluated and monitored. The practice of using bibliometrics in these processes has become widely accepted around the world, and therefore scientists often try to adapt their publishing practices to achieve high bibliometric indicators. Unfortunately, in practice, in the absence of a qualitative analysis of scientific results, such a focus on bibliometric indicators frequently leads to abuse and violations.

Changes in the official requirements for the evaluation of research results in Ukraine, especially for the accreditation of universities and procedures for obtaining academic degrees and ranks, have encouraged Ukrainian authors to publish their research results primarily in foreign journals and/or in journal titles that are indexed in WoS and Scopus. Unfortunately, an increase in the number of these publications is accompanied by an increase in publications by Ukrainian authors in Scopus-delisted titles, some of which are openly questionable titles. The current Ukrainian legislation and evaluation methods of academics do not warn authors against such journals.

In most subject fields, papers in Scopus-delisted titles comprise a small percentage of the total number of publications in Ukraine, but not for all. The situation is particularly worrying in the field of economics, econometrics and finance, where almost every second Ukrainian paper appears in a Scopus-delisted journal. At the same time, the majority of doctor and PhD diplomas in Ukraine are in economic sciences (Roman Radejko's blog 2021). Thus, there is a risk that for many Ukrainian authors in the field of economics, inclusion of their paper in a Scopus journal meets formal requirements more than presenting their research results to an international audience.

The analysis of papers of Ukrainian authors in Scopus-delisted journal titles based on institutions indicates that in dozens of Ukrainian institutions, such papers make up a significant portion of all those institutions' publications, and in the case of three institutions, these papers accounted for almost half of all their publications. Similarly, the analysis of citations of papers in Scopus-delisted journal titles indicates that in several Ukrainian institutions, the citation of such papers significantly exceeds the expected number of citations in comparison with other papers of the same year and in the same field. All these facts may indicate that some authors use the shortcomings of Ukrainian law to obtain undeserved rewards. Therefore, experts should pay special attention to such facts when evaluating the scientific results of Ukrainian academics and institutions.

A very high percentage of publications that were detected in Scopus-delisted economics, econometrics and finance journals were from three countries of the former Soviet Union, Ukraine, Russia and Belarus, whose long-term economic ideology developed into a "planned economy" (Harrison, 2017). Thus, the high percentage of such publications in this area may signal a weakness of economic research in these countries, due to their historical features. Therefore, researchers in these countries and in these subject areas need a special curricular development plan and more time to gain the necessary experience and to build a reliable network of scientific contacts. The use of "common rules" for all fields, conversely, can provoke an increase in such "problems" areas and increase the number of papers in questionable titles.

It is possible that some Ukrainian authors have created a deliberate symbiosis with dubious publishers, but it is also likely that many authors have chosen dubious journals to publish their results because of their general low information culture. The emergence of new state requirements for the publication of research results in "international journals" in 2012 led primarily to an increase in the number of publications by Ukrainian authors in journals of Ukraine's neighbors, in particular in journals of the Russian Federation (Nazarovets & Nazarovets, 2018), as could be predicted. The ability to publish articles in Russian is an important option for Ukrainian authors, especially in the social sciences and humanities, as they write fewer scientific papers in English than authors in the natural sciences. Accordingly, the changes made by the Ukrainian government to the national scientific policy at that time, especially without increasing funding for science, did not lead to more Ukrainians publishing many more articles in leading international journals.

The occupation of the Crimean Peninsula by the armed forces of the Russian Federation in 2014 and the subsequent Russian armed aggression in eastern Ukraine, which continues to this day, has certainly deterred Ukrainian authors from publishing their work in Russian journals (Nazarovets, 2017). Thus, many Ukrainian academics had to quickly look for new channels to disseminate their scientific results. Consequently, those authors with no experience of publishing in influential international journals could become easy prey for aggressive marketing of questionable publishers. This surge of Ukrainian papers in questionable titles was predictable, but Ukrainian officials have systematically ignored – and continue to ignore – this important problem. The papers in Scopus-delisted titles, as well as citations of these papers, continue to be considered and rewarded on par with other papers and their citations. As a result, the number of such papers continues to grow and this creates a critical situation in some scientific fields and institutions of Ukraine.

This study focused on an analysis of publications in Scopus-delisted titles by subject fields, rather than at the individual level, because the issue of increasing the number of such papers is primarily a concern of shortcomings in the Ukrainian evaluation system, rather than the behavior of individual researchers. Furthermore, authors could have submitted their manuscripts in good faith to Scopus-indexed titles in order to meet government and institutional requirements in the hope that those publications would benefit their scientific careers. Due to pressure from government and institutional requirements, authors are usually interested in publishing their results as soon as possible. Therefore, it is possible some authors could immediately focus on Scopus journal titles that often publish new papers without taking into account other characteristics of those titles. Therefore, some authors could publish quality papers in delisted titles and some of these papers could be peer-reviewed.

It is obvious that Ukrainian legislative initiatives aimed at improving the quality of national research and their better presentation are partially depreciated by the incorrect use of bibliometric indicators. Therefore, Ukrainian scientific officials need to take decisive steps to combat the practice of publishing work in questionable titles. First of all, it is necessary to make changes in all accepted national methods of evaluation of scientific results in order to make it impossible to equate papers in questionable titles with papers in reputable titles.

According to the results, Hungary and Belarus have the lowest percentage of publications in Scopus-delisted titles relative to their total number of publications. In my opinion, this does not imply that the evaluation systems of these countries provide safeguards against the practice of publishing in questionable journal titles, but rather suggests that the spread of this practice may be influenced by various factors. For example, in Belarus, there are no additional rewards for authors who publish articles in Scopus-indexed journals, so Belarusian authors are likely to find it more convenient to comply with government requirements by publishing in national journals.

At the same time, authors' level of information culture and their knowledge of best practices of scientific communication is extremely important. Researchers must have universal criteria for evaluating publications, which will allow them to better navigate whether they can be trusted to present their research output. An example of such criteria is the *Think. Check. Submit.* checklist that helps researchers identify trusted journals and publishers[5]. In addition, there are independent web tools that help authors identify trusted journals, including those that use crowdsourcing (van Gerestein, 2015). In international collaborations, Ukrainian authors have published significantly fewer papers in Scopus-delisted titles, so cooperation and exchange of best practices between researchers from different countries also contribute to a better choice of communication channels for the dissemination of research output.

[5] https://thinkchecksubmit.org/

It is also necessary to conduct an expert inspection of work by authors and institutions, especially those who are particularly actively published in Scopus-delisted journals, to identify the causes of such behavior and to try and eliminate these shortcomings. Moreover, experts need to carefully check papers for the possible manipulation of citations. In contrast to the problem of selecting questionable journals, in which it can be assumed that an author's guilt may be minimal, the situation with “suspicious” citations is considerably different. If authors regularly manipulate citations to increase their bibliometric indicators, it can no longer be said that this is only the shortcoming of the evaluation system, because the authors deliberately and purposefully resort to fraud. Some Ukrainian scientists are world leaders in extreme self-citation (Ioannidis et al., 2019), and the fact that conference papers in Scopus-delisted journal titles are cited more often than regular articles is alarming because conference proceedings are usually rarely cited (Michels & Fu, 2014).

Of course, it is very difficult to develop effective approaches to the correct consideration of citations, but here it is possible to try and apply some new proposals that might improve the country’s scientific practices (Teixeira da Silva & Vuong, 2021). For example, when applying for research grants, researchers may notify the organizers of the competition and refuse suspicious citations. Just as any researcher can remove irrelevant publications from their Scopus profile, researchers should be able to voluntarily report possible irregularities in citations to their publications in order to hold fair competitions for research projects. Moreover, an effort should be made to balance the weight of indicators that take into account the number of citations, such as the h-index, with other indicators of evaluation, such as the Global Science Factor (Teixeira da Silva, 2013). Accordingly, the lack of significant motivation of authors to artificially increase their number of citations can also reduce the number of publications in questionable journal titles. However, in parallel, penalties should be introduced for authors and institutions that systematically publish works in questionable journals, as has been done successfully, for example, in China (Cyranoski, 2018).

Another promising step that could prevent authors from falsifying their scientific achievements is the introduction of guidelines for verifiable, accurate, complete, updated, and public *curriculum vitae* at the national level (Teixeira da Silva et al., 2020). It is also necessary to conduct educational activities on the dangers of publications in questionable journals, especially among those Ukrainian authors working in fields where the practice of such publications is particularly common. Identifying and eliminating shortcomings in the Ukrainian evaluation system will help to shape Ukraine's best science policy in the long run.

This study was conducted in peacetime. The results of this work were intended to help policymakers in different Eastern European countries to improve national research evaluation systems for human progress. Russia, with Belarusian support, launched a full-scale military invasion of Ukraine on February 24, 2022. This Russian invasion of Ukraine is the largest military attack on a sovereign state in Europe since the World War II. Of course, now in Ukraine, the problem of the correct use of bibliometrics in the evaluation system is out-of-sync with this wartime reality. However, in a hopefully future time of peace, Ukraine will always need a strong system of education, science, and publishing culture, so the results of this study should help to avoid similar mistakes in the evaluation of research outputs in the future.

*Acknowledgements*

The author thanks all the defenders of Ukraine, whose desperate struggle against the occupiers allowed the author to complete this study. The author thanks Maryna Nazarovets and Jaime A. Teixeira da Silva for their support and constructive feedback. Jaime A. Teixeira da Silva also text-edited the paper and conducted an English revision. Special thanks to the editor, Pipa Smart, and four anonymous referees for numerous excellent suggestions that greatly improved this paper.

And just in case - Russian warship, go f**k yourself!

*Declarations*

The author has no conflicts of interest to declare that are relevant to the content of this article. This research received no specific grant from any funding agency in the public, commercial, or not-for-profit sectors.

**Table 1** Research and development expenditure (% of GDP) in selected countries of Eastern Europe in 2011-2018

| Country | 2011 | 2012 | 2013 | 2014 | 2015 | 2016 | 2017 | 2018 |
|---|---|---|---|---|---|---|---|---|
| Czech Republic | 1,56 | 1,78 | 1,90 | 1,97 | 1,93 | 1,68 | 1,79 | 1,93 |
| Hungary | 1,19 | 1,26 | 1,39 | 1,35 | 1,35 | 1,19 | 1,33 | 1,53 |
| Poland | 0,75 | 0,88 | 0,87 | 0,94 | 1,00 | 0,96 | 1,03 | 1,21 |
| Russian Federation | 1,02 | 1,03 | 1,03 | 1,07 | 1,10 | 1,10 | 1,11 | 0,98 |
| Slovak Republic | 0,66 | 0,80 | 0,82 | 0,88 | 1,16 | 0,79 | 0,89 | 0,84 |
| Belarus | 0,68 | 0,65 | 0,65 | 0,51 | 0,50 | 0,50 | 0,58 | 0,60 |
| Romania | 0,50 | 0,48 | 0,39 | 0,38 | 0,49 | 0,48 | 0,50 | 0,50 |
| Ukraine | 0,74 | 0,75 | 0,76 | 0,65 | 0,61 | 0,48 | 0,45 | 0,47 |

**Table 2** Publication activity of Ukrainian authors in Scopus-delisted journal titles in which they have published the most papers in 2011-2020

| Journal titles | 2011 | 2012 | 2013 | 2014 | 2015 | 2016 | 2017 | 2018 | 2019 | 2020 | Total |
|---|---|---|---|---|---|---|---|---|---|---|---|
| *Actual Problems of Economics* | 10 | 234 | 301 | 410 | 398 | 322 | x | x | x | x | **1675** |
| *Advances in Intelligent Systems and Computing* | x | 0 | 2 | 7 | 16 | 13 | 69 | 68 | 173 | 337 | **685** |
| *IOP Conference Series Materials Science and Engineering* | 3 | 5 | 1 | 4 | 17 | 22 | 70 | 32 | 204 | 227 | **585** |
| *Metallurgical and Mining Industry* | 62 | 32 | 19 | 82 | 222 | x | x | x | x | x | **417** |
| *MATEC Web of Conferences* | x | 0 | 1 | 1 | 7 | 17 | 88 | 160 | 0 | x | **274** |
| *Journal of Water Chemistry and Technology* | 47 | 29 | 35 | 36 | 46 | 35 | 44 | x | x | x | **272** |
| *Journal of Advanced Research in Law and Economics* | 0 | 0 | 1 | 0 | 0 | 5 | 23 | 49 | 91 | 36 | **205** |
| *International Journal of Engineering and Technology (UAE)* | x | x | x | x | x | 2 | 0 | 188 | x | x | **190** |
| *EPJ Web of Conferences* | 2 | 8 | 15 | 16 | 8 | 29 | 18 | 34 | x | x | **130** |
| *Asia Life Sciences* | 0 | 0 | x | x | x | 0 | 0 | 0 | 62 | 44 | **106** |

x – this title was not indexed in Scopus in this year

**Table 3** Papers of Ukrainian authors for 2011-2020 in Scopus-delisted titles based on the amount of international, national and institutional collaborations

| Types of collaboration | Percentage | Papers | Citations | Citations per Paper | Field-Weighted Citation Impact |
|---|---|---|---|---|---|
| International | 18.4% | 1267 | 5857 | 4.6 | 1.43 |
| Only national | 27.7% | 1841 | 5597 | 3.0 | 0.86 |
| Only institutional | 33.6% | 2317 | 6097 | 2.6 | 0.79 |
| Single authorship | 21.4% | 1477 | 1511 | 1.0 | 0.15 |

**Table 4** Papers of Ukrainian authors for 2011-2020 in Scopus-delisted titles by subject field and the percentage of total number of Ukrainian papers in each field

| Subject Field | Number of Papers in Delisted Journal Titles | Percentage of Papers in Delisted Journal Titles | Number of Delisted Journal Titles | Citations | Number of Authors | Citations per Publication in Delisted Journal Titles | Field-Weighted Citation Impact |
|---|---|---|---|---|---|---|---|
| Economics, Econometrics and Finance | 2216 | 46.92% | 25 | 3108 | 3551 | 1.4 | 0.22 |
| Pharmacology, Toxicology and Pharmaceutics | 376 | 18.60% | 32 | 848 | 1170 | 2.3 | 0.42 |
| Multidisciplinary | 80 | 11.90% | 6 | 218 | 156 | 2.7 | 0.09 |
| Environmental Science | 701 | 11.79% | 25 | 1952 | 1709 | 2.8 | 0.33 |
| Social Sciences | 765 | 11.65% | 36 | 1927 | 2496 | 2.5 | 0.59 |
| Business, Management and Accounting | 537 | 10.41% | 29 | 1867 | 1979 | 3.5 | 0.83 |
| Engineering | 3042 | 8.70% | 58 | 11330 | 7633 | 3.7 | 1.41 |
| Biochemistry, Genetics and Molecular Biology | 481 | 6.58% | 29 | 1185 | 1627 | 2.5 | 0.34 |
| Computer Science | 1316 | 6.10% | 27 | 6129 | 3617 | 4.7 | 1.93 |
| Earth and Planetary Sciences | 449 | 5.93% | 5 | 1658 | 860 | 3.7 | 0.26 |
| Chemical Engineering | 289 | 5.50% | 11 | 819 | 960 | 2.8 | 0.31 |
| Decision Sciences | 191 | 5.48% | 6 | 471 | 826 | 2.5 | 0.56 |
| Chemistry | 636 | 4.91% | 14 | 2183 | 1457 | 3.4 | 0.96 |
| Materials Science | 1341 | 4.90% | 16 | 4225 | 3216 | 3.2 | 0.94 |
| Arts and Humanities | 91 | 3.62% | 13 | 102 | 256 | 1.1 | 0.45 |
| Agricultural and Biological Sciences | 184 | 3.20% | 18 | 670 | 625 | 3.6 | 0.39 |
| Veterinary | 3 | 1.65% | 1 | 10 | 12 | 3.3 | 0.76 |
| Energy | 138 | 1.41% | 5 | 555 | 540 | 4 | 0.58 |
| Neuroscience | 9 | 1.17% | 1 | 20 | 20 | 2.2 | 0.07 |
| Health Professions | 8 | 0.77% | 1 | 12 | 32 | 1.5 | 0.2 |
| Mathematics | 102 | 0.63% | 20 | 695 | 235 | 6.8 | 0.72 |
| Medicine | 59 | 0.60% | 20 | 205 | 191 | 3.5 | 0.34 |
| Physics and Astronomy | 201 | 0.53% | 8 | 504 | 2149 | 2.5 | 0.61 |
| Psychology | 2 | 0.37% | 2 | 0 | 7 | 0 | 0 |
| Immunology and Microbiology | 2 | 0.22% | 1 | 24 | 22 | 12 | 0.82 |

**Table 5** Papers of Ukrainian authors for 2011-2020 in Scopus-delisted titles by institutions and the percentage of total number of Ukrainian papers in each field

| Institution | Total Number of Papers | Number of Papers in Delisted Journal Titles | Percent age of Papers in Delisted Titles | Citations | Author s | Citation s per Paper in Delisted Journal Titles | Field-W eighted Citation Impact |
|---|---|---|---|---|---|---|---|
| National Academy of Sciences of Ukraine** | 34320 | 766 | 2.23% | 2006 | 1006 | 2.6 | 0.64 |
| Lviv Polytechnic National University | 5739 | 447 | 7.79% | 2806 | 561 | 6.3 | 2.9 |
| Kyiv National Taras Shevchenko University | 10901 | 416 | 3.82% | 898 | 566 | 2.2 | 0.65 |
| Sumy State University | 2545 | 254 | 9.98% | 1176 | 292 | 4.6 | 0.67 |
| National Technical University of Ukraine "Igor Sikorsky Kyiv Polytechnic Institute" | 5457 | 242 | 4.43% | 777 | 322 | 3.2 | 1.29 |
| Dumansky Institute of Colloid and Water Chemistry | 377 | 198 | 52.52% | 410 | 175 | 2.1 | 0.11 |
| National Aviation University | 1928 | 185 | 9.60% | 589 | 221 | 3.2 | 1.31 |
| Poltava National Technical Yuriy Kondratyuk University | 420 | 182 | 43.33% | 469 | 245 | 2.6 | 0.49 |
| National University of Life and Environmental Sciences of Ukraine | 1399 | 175 | 12.51% | 529 | 190 | 3 | 0.86 |
| Kyiv National University of Trade and Economics | 546 | 163 | 29.85% | 410 | 186 | 2.5 | 0.49 |
| Kryvyi Rih National University | 696 | 162 | 23.28% | 926 | 135 | 5.7 | 0.49 |
| National University of Pharmacy | 768 | 155 | 20.18% | 189 | 241 | 1.2 | 0.13 |
| Kyiv National Economic University named after Vadym Hetman | 401 | 133 | 33.17% | 172 | 153 | 1.3 | 0.29 |
| Kharkiv National University of Radio Electronics | 2218 | 122 | 5.50% | 733 | 119 | 6 | 1.62 |
| Ivan Franko National University of L'viv | 3631 | 116 | 3.19% | 284 | 142 | 2.4 | 0.95 |
| Ukrainian State University of Railway Transport | 466 | 114 | 24.46% | 320 | 163 | 2.8 | 1.15 |
| Kyiv National University of Construction and Architecture | 506 | 111 | 21.94% | 363 | 156 | 3.3 | 1.43 |
| V. N. Karazin Kharkiv National University | 5392 | 110 | 2.04% | 281 | 123 | 2.6 | 0.78 |
| National Metallurgical Academy of Ukraine | 498 | 102 | 20.48% | 119 | 139 | 1.2 | 0.19 |
| Yaroslav Mudryi National Law University | 310 | 94 | 30.32% | 203 | 130 | 2.2 | 0.63 |
| Kyiv National University of Technologies and Design | 463 | 94 | 20.30% | 142 | 106 | 1.5 | 0.32 |
| Lutsk National Technical University | 444 | 87 | 19.59% | 155 | 82 | 1.8 | 0.76 |
| Vasyl Stefanyk Precarpathian National University | 917 | 86 | 9.38% | 297 | 84 | 3.5 | 1.23 |
| Simon Kuznets Kharkiv National University of Economics | 434 | 84 | 19.35% | 246 | 115 | 2.9 | 0.56 |
| National University of Water and Environmental Engineering | 584 | 84 | 14.38% | 322 | 93 | 3.8 | 1.68 |
| O.M. Beketov National University of Urban Economy in Kharkiv | 473 | 80 | 16.91% | 114 | 113 | 1.4 | 0.55 |
| National Aerospace University "Kharkiv Aviation Institute" | 1197 | 80 | 6.68% | 271 | 129 | 3.4 | 1.96 |
| Oles Honchar Dnipropetrovsk National University | 1731 | 80 | 4.62% | 242 | 95 | 3 | 0.65 |
| West Ukrainian National University | 696 | 76 | 10.92% | 214 | 86 | 2.8 | 0.81 |
| State University of Infrastructure and Technologies | 260 | 71 | 27.31% | 273 | 73 | 3.8 | 1.11 |
| National Technical University Kharkiv Polytechnic Institute | 2616 | 71 | 2.71% | 298 | 113 | 4.2 | 1.34 |
| Ministry of Education and Science of Ukraine** | 1755 | 69 | 3.93% | 233 | 97 | 3.4 | 0.74 |
| Pryazovskyi State Technical University | 399 | 67 | 16.79% | 375 | 65 | 5.6 | 1.64 |
| University of the State Fiscal Service of Ukraine | 187 | 63 | 33.69% | 158 | 84 | 2.5 | 0.54 |
| Odessa I.I.Mechnikov National University | 1531 | 63 | 4.11% | 204 | 69 | 3.2 | 0.5 |
| Sumy National Agrarian University | 392 | 61 | 15.56% | 225 | 95 | 3.7 | 0.65 |
| Kremenchuk Mykhailo Ostrohradskyi National University | 549 | 61 | 11.11% | 262 | 73 | 4.3 | 0.46 |
| Khmelnytsky National University | 549 | 61 | 11.11% | 161 | 71 | 2.6 | 0.78 |
| Ivano-Frankivsk National Technical University of Oil and Gas | 766 | 60 | 7.83% | 231 | 76 | 3.8 | 1.64 |
| Volodymyr Dahl East Ukrainian National University | 584 | 58 | 9.93% | 282 | 85 | 4.9 | 1.42 |
| Odessa National Polytechnic University | 717 | 58 | 8.09% | 364 | 73 | 6.3 | 1.11 |
| Ivan Kozhedub Kharkiv National Air Force University | 412 | 56 | 13.59% | 588 | 84 | 10.5 | 3.19 |
| Odessa National Academy of Food Technologies | 474 | 56 | 11.81% | 321 | 50 | 5.7 | 1.09 |
| Kharkiv Petro Vasylenko National Technical University of Agriculture | 345 | 55 | 15.94% | 295 | 65 | 5.4 | 1.35 |
| Zaporizhzhya National University | 527 | 53 | 10.06% | 130 | 75 | 2.5 | 0.53 |
| Zaporizhia National Technical University | 665 | 53 | 7.97% | 165 | 69 | 3.1 | 0.64 |
| Institute of Semiconductors Physics | 2080 | 52 | 2.50% | 153 | 94 | 2.9 | 0.99 |
| Mykhailo Tuhan-Baranovskyi Donetsk National University of Economics and Trade | 182 | 51 | 28.02% | 39 | 53 | 0.8 | 0.15 |
| Lesya Ukrainka Eastern European National University | 610 | 51 | 8.36% | 115 | 60 | 2.3 | 0.66 |
| National Bank of Ukraine | 104 | 49 | 47.12% | 173 | 58 | 3.5 | 0.33 |

**The table shows the first 50 institutions by number of papers in Scopus-delisted journal titles*
***Under this profile, Scopus combines several institutions*

**Table 6** Number and percentage of papers in Scopus-delisted journal titles of authors from seven Eastern European countries in 2011-2020. Top-3 subject areas based on the percentage of papers in Scopus-delisted titles by countries

| Countries | Number of publications in Scopus-delisted titles | Percentage of publications in Scopus-delisted titles |
|---|---|---|
| **Belarus** | 425 | 2.11% |
| Business, Management and Accounting | 21 | 15.44% |
| Economics, Econometrics and Finance | 16 | 14.68% |
| Social Sciences | 34 | 6.48% |
| **Czech Republic** | 7436 | 3.30% |
| Engineering | 4098 | 8.86% |
| Economics, Econometrics and Finance | 268 | 5.53% |
| Materials Science | 1555 | 5.37% |
| **Hungary** | 1211 | 1.14% |
| Materials Science | 436 | 5.25% |
| Engineering | 686 | 4.62% |
| Pharmacology, Toxicology and Pharmaceutics | 117 | 2.79% |
| **Poland** | 12382 | 2.88% |
| Engineering | 9209 | 10.59% |
| Economics, Econometrics and Finance | 557 | 9.33% |
| Computer Science | 4277 | 8.43% |
| **Romania** | 13053 | 8.70% |
| Veterinary | 208 | 33.66% |
| Materials Science | 8334 | 31.07% |
| Chemistry | 5626 | 28.49% |
| **Russian Federation** | 54678 | 7.09% |
| Economics, Econometrics and Finance | 4912 | 31.49% |
| Business, Management and Accounting | 4350 | 28.18% |
| Pharmacology, Toxicology and Pharmaceutics | 3395 | 27.82% |
| **Ukraine** | 6902 | 5.72% |
| Economics, Econometrics and Finance | 4723 | 46.92% |
| Pharmacology, Toxicology and Pharmaceutics | 2021 | 18.60% |
| Environmental Science | 5944 | 11.79% |